\documentclass[aip,jcp,reprint,noshowkeys,superscriptaddress]{revtex4-1}
\usepackage{graphicx,dcolumn,bm,xcolor,microtype,multirow,amsmath,amssymb,amsfonts,physics,mhchem}

\usepackage[utf8]{inputenc}
\usepackage[T1]{fontenc}
\usepackage{txfonts}
\usepackage{grffile}

\usepackage[
	colorlinks=true,
    citecolor=blue,
    breaklinks=true
	]{hyperref}
\usepackage[normalem]{ulem}

\newcommand{\ie}{\textit{i.e.}}
\newcommand{\eg}{\textit{e.g.}}

\newcommand{\mc}{\multicolumn}

\newcommand{\fnm}{\footnotemark}
\newcommand{\fnt}{\footnotetext}
\newcommand{\tabc}[1]{\multicolumn{1}{c}{#1}}

\newcommand{\cD}{\mathcal{D}}

\newcommand{\hH}{\Hat{H}}
\newcommand{\hHc}{\Hat{h}}
\newcommand{\hT}{\Hat{T}}
\newcommand{\hWee}{\Hat{W}_\text{ee}}
\newcommand{\hGam}[1]{\Hat{\Gamma}^{#1}}
\newcommand{\hgam}[1]{\Hat{\gamma}^{#1}}
\newcommand{\hgamdens}[1]{\Hat{\gamma}^{#1}[n]}

\newcommand{\hVne}{\Hat{V}_\text{ne}}
\newcommand{\vne}{v_\text{ne}}

\newcommand{\F}[2]{F_{#1}^{#2}}
\newcommand{\Ts}[1]{T_\text{s}^{#1}}
\newcommand{\eps}[2]{\varepsilon_{#1}^{#2}}
\newcommand{\Eps}[2]{\mathcal{E}_{#1}^{#2}}
\newcommand{\e}[2]{\epsilon_{#1}^{#2}}

\newcommand{\E}[2]{E_{#1}^{#2}}

\newcommand{\be}[2]{\overline{\epsilon}_{#1}^{#2}}
\newcommand{\n}[2]{n_{#1}^{#2}}
\newcommand{\Cx}[1]{C_\text{x}^{#1}}

\newcommand{\HF}{\text{HF}}

\newcommand{\VWN}{\text{VWN5}}
\newcommand{\eVWN}{\text{eVWN5}}

\newcommand{\LIM}{\text{LIM}}
\newcommand{\MOM}{\text{MOM}}
\newcommand{\Hxc}{\text{Hxc}}
\newcommand{\Ha}{\text{H}}
\newcommand{\ex}{\text{x}}
\newcommand{\co}{\text{c}}
\newcommand{\xc}{\text{xc}}

\newcommand{\br}{\boldsymbol{r}}
\newcommand{\bw}{\boldsymbol{w}}

\newcommand{\Ex}[2]{\Omega_{#1}^{#2}}

\newcommand{\ew}[1]{w_{#1}}

\newcommand{\ON}[2]{f_{#1}^{#2}}
\newcommand{\Det}[2]{\Phi_{#1}^{#2}}

\newcommand{\nEns}{M}

\newcommand{\nOrb}{K}

\newcommand{\MO}[2]{\phi_{#1}^{#2}}

\newcommand{\RHH}{R_{\ce{H-H}}}

\newcommand{\LCPQ}{Laboratoire de Chimie et Physique Quantiques, Universit\'e de Toulouse, CNRS, UPS, France}
\newcommand{\LCQ}{Laboratoire de Chimie Quantique, Institut de Chimie, CNRS, Universit\'e de Strasbourg, Strasbourg, France}
\newcommand{\UL}{Instituut-Lorentz, Universiteit Leiden, P.O.~Box 9506, 2300 RA Leiden, The Netherlands}
\newcommand{\VU}{Division of Theoretical Chemistry, Vrije Universiteit Amsterdam, De Boelelaan 1083, 1081 HV Amsterdam, The Netherlands}

\newcommand{\beq}{\begin{eqnarray}}
\newcommand{\eeq}{\end{eqnarray}}

\begin{document}	

\title{Weight Dependence of Local Exchange-Correlation Functionals in Ensemble Density-Functional Theory: Double Excitations in Two-Electron Systems}

\author{Clotilde \surname{Marut}}
	\affiliation{\LCPQ}
\author{Bruno \surname{Senjean}}
	\affiliation{\UL}
	\affiliation{\VU}
\author{Emmanuel \surname{Fromager}}
	\affiliation{\LCQ}
\author{Pierre-Fran\c{c}ois \surname{Loos}}
	\email{loos@irsamc.ups-tlse.fr}
	\affiliation{\LCPQ}

\begin{abstract}
Gross--Oliveira--Kohn (GOK) ensemble density-functional theory (GOK-DFT)
is a time-\textit{independent} extension of density-functional theory (DFT) which
allows to compute excited-state
energies via the derivatives of the ensemble energy with
respect to the ensemble weights.
Contrary to the time-dependent version of DFT (TD-DFT), double excitations can be easily computed within GOK-DFT.
However, to take full advantage of this formalism, one must have access to a \textit{weight-dependent} exchange-correlation functional in order to model the infamous ensemble derivative contribution to the excitation energies.
In the present article, we discuss the construction of first-rung (\textit{i.e.}, local) weight-dependent exchange-correlation density-functional approximations for two-electron atomic and molecular systems (He and H$_2$) specifically designed for the computation of double excitations within GOK-DFT.
In the spirit of optimally-tuned range-separated hybrid functionals, a two-step system-dependent procedure is proposed to obtain accurate energies associated with double excitations.
\end{abstract}

\maketitle

\section{Introduction}
Time-dependent density-functional theory (TD-DFT) has been the dominant force in the calculation of excitation energies of molecular systems in the last two decades.\cite{Casida_1995,Ulrich_2012,Loos_2020a}
At a moderate computational cost (at least compared to the other excited-state \textit{ab initio} methods), TD-DFT can provide accurate transition energies for low-lying excited states of organic molecules (see, for example, Ref.~\onlinecite{Dreuw_2005} and references therein).
Importantly, within the widely-used adiabatic approximation, setting up a TD-DFT calculation for a given system is an
almost pain-free process from a user perspective as the only (yet
essential) input variable is the choice of the 
ground-state exchange-correlation (xc) functional.

Similar to density-functional theory (DFT), \cite{Hohenberg_1964,Kohn_1965,ParrBook} TD-DFT is an in-principle exact theory which formal foundations rely on the Runge-Gross theorem. \cite{Runge_1984}
The Kohn-Sham (KS) formulation of TD-DFT transfers the
complexity of the many-body problem to the xc functional thanks to a
judicious mapping between a time-dependent non-interacting reference
system and its interacting analog which both have
exactly the same one-electron density.

However, TD-DFT is far from being perfect as, in practice, drastic approximations must be made.
First, within the commonly used linear-response regime, the electronic spectrum relies on the (unperturbed) pure-ground-state KS picture, \cite{Runge_1984, Casida_1995, Casida_2012} which may not be adequate in certain situations (such as strong correlation). 
Second, the time dependence of the functional is usually treated at the local approximation level within the standard adiabatic approximation.
In other words, memory effects are absent from the xc functional which is assumed to be local in time
(the xc energy is in fact an xc action, not an energy functional). \cite{Vignale_2008}
Third and more importantly in the present context, a major issue of
TD-DFT actually originates directly from the choice of the (ground-state) xc functional, and more specifically, the possible (not to say likely) substantial variations in the quality of the excitation energies for two different choices of xc functionals.

Because of its popularity, approximate TD-DFT has been studied extensively, and some researchers have quickly unveiled various theoretical and practical deficiencies.
For example, TD-DFT has problems with charge-transfer \cite{Tozer_1999,Dreuw_2003,Sobolewski_2003,Dreuw_2004,Maitra_2017} and Rydberg \cite{Tozer_1998,Tozer_2000,Casida_1998,Casida_2000,Tozer_2003} excited states (the excitation energies are usually drastically underestimated) due to the wrong asymptotic behaviour of the semi-local xc functional.
The development of range-separated hybrids provides an effective solution to this problem. \cite{Tawada_2004,Yanai_2004} 
From a practical point of view, the TD-DFT xc kernel is usually considered as static instead of being frequency dependent.
One key consequence of this so-called adiabatic approximation (based on the assumption that the density varies slowly with time) is that double excitations are completely absent from the TD-DFT spectra. \cite{Levine_2006,Tozer_2000,Elliott_2011}
Although these double excitations are usually experimentally dark (which means that they usually cannot be observed in photo-absorption spectroscopy), these states play, indirectly, a key role in many photochemistry mechanisms. \cite{Boggio-Pasqua_2007} They are, moreover, a real challenge for high-level computational methods. \cite{Loos_2018,Loos_2019,Loos_2020b}

One possible solution to access double excitations within TD-DFT is provided by spin-flip TD-DFT which describes double excitations as single excitations from the lowest triplet state. \cite{Huix-Rotllant_2010,Krylov_2001,Shao_2003,Wang_2004,Wang_2006,Minezawa_2009}
However, spin contamination might be an issue. \cite{Huix-Rotllant_2010}
Note that a simple remedy based on a mixed reference reduced density
matrix has been recently introduced by Lee \textit{ et al.} \cite{Lee_2018}
In order to go beyond the adiabatic approximation, a dressed TD-DFT approach has been proposed by Maitra and coworkers \cite{Maitra_2004,Cave_2004} (see also Refs.~\onlinecite{Mazur_2009,Mazur_2011,Huix-Rotllant_2011,Elliott_2011,Maitra_2012}).
In this approach the xc kernel is made frequency dependent, which allows to treat doubly-excited states. \cite{Romaniello_2009a,Sangalli_2011,Loos_2019}

Maybe surprisingly, another possible way of accessing double excitations is to resort to a time-\textit{independent} formalism. \cite{Yang_2017,Sagredo_2018,Deur_2019}
With a computational cost similar to traditional KS-DFT, DFT for
ensembles (eDFT)
\cite{Theophilou_1979,Gross_1988a,Gross_1988b,Oliveira_1988} is a viable
alternative that follows such a strategy and is currently under active development.\cite{Gidopoulos_2002,Franck_2014,Borgoo_2015,Kazaryan_2008,Gould_2013,Gould_2014,Filatov_2015,Filatov_2015b,Filatov_2015c,Gould_2017,Deur_2017,Gould_2018,Gould_2019a,Sagredo_2018,Ayers_2018,Deur_2018,Deur_2019,Kraisler_2013,Kraisler_2014,Alam_2016,Alam_2017,Nagy_1998,Nagy_2001,Nagy_2005,Pastorczak_2013,Pastorczak_2014,Pribram-Jones_2014,Yang_2013a,Yang_2014,Yang_2017,Senjean_2015,Senjean_2016,Smith_2016,Senjean_2018}
In the assumption of monotonically decreasing weights, eDFT for excited states has the undeniable advantage to be based on a rigorous variational principle for ground and excited states,  the so-called Gross--Oliveria--Kohn (GOK) variational principle.  \cite{Gross_1988a} 
In short, GOK-DFT (\ie, eDFT for neutral excitations) is the density-based analog of state-averaged wave function methods, and excitation energies can then be easily extracted from the total ensemble energy. \cite{Deur_2019}
Although the formal foundations of GOK-DFT have been set three decades ago, \cite{Gross_1988a,Gross_1988b,Oliveira_1988} its practical developments have been rather slow.
We believe that it is partly due to the lack of accurate approximations for GOK-DFT.
In particular, to the best of our knowledge, although several attempts have been made, \cite{Nagy_1996,Paragi_2001} an explicitly
weight-dependent density-functional approximation for ensembles (eDFA)
has never been developed for atoms and molecules from first principles.
The present contribution paves the way towards this goal.

The local-density approximation (LDA), as we know it, is based on the uniform electron gas (UEG) also known as jellium, an hypothetical infinite substance where an infinite number of electrons ``bathe'' in a (uniform) positively-charged jelly. \cite{Loos_2016}
Although the Hohenberg--Kohn theorems \cite{Hohenberg_1964} are here to provide firm theoretical grounds to DFT, modern KS-DFT rests largely on the presumed similarity between this hypothetical UEG and the electronic behaviour in a real system. \cite{Kohn_1965}
However, Loos and Gill have recently shown that there exists other UEGs which contain finite numbers of electrons (more like in a molecule), \cite{Loos_2011b,Gill_2012} and that they can be exploited to construct ground-state functionals as shown in Refs.~\onlinecite{Loos_2014a,Loos_2014b,Loos_2017a}, where the authors proposed generalised LDA exchange and correlation functionals.

Electrons restricted to remain on the surface of a $\cD$-sphere (where $\cD$ is the dimensionality of the surface of the sphere) are an example of finite UEGs (FUEGs). \cite{Loos_2011b}
Very recently, \cite{Loos_2020} two of the present authors have taken advantages of these FUEGs to construct a local, weight-dependent correlation functional specifically designed for one-dimensional many-electron systems.
Unlike any standard functional, this first-rung functional automatically incorporates ensemble derivative contributions thanks to its natural weight dependence,  \cite{Levy_1995, Perdew_1983} and has shown to deliver accurate excitation energies for both single and double excitations.
In order to extend this methodology to more realistic (atomic and molecular) systems, we combine here these FUEGs with the usual infinite UEG (IUEG) to construct a weigh-dependent LDA correlation functional for ensembles, which is specifically designed to compute double excitations within GOK-DFT.

The paper is organised as follows.
In Sec.~\ref{sec:theo}, the theory behind GOK-DFT is briefly presented.
Section \ref{sec:compdet} provides the computational details.
The results of our calculations for two-electron systems are reported and discussed in Sec.~\ref{sec:res}. 
Finally, we draw our conclusions in Sec.~\ref{sec:ccl}. 
Unless otherwise stated, atomic units are used throughout.

\section{Theory}
\label{sec:theo}

Let us consider a GOK ensemble of $\nEns$ electronic states with
individual energies $\E{}{(0)} \le \ldots \le \E{}{(\nEns-1)}$, and
(normalised) monotonically decreasing weights $\bw \equiv (\ew{1},\ldots,\ew{M-1})$, \ie, $\ew{0}=1-\sum_{I=1}^{\nEns-1} \ew{I}$, and $\ew{0} \ge \ldots \ge \ew{\nEns-1}$.
The corresponding ensemble energy
\begin{equation}\label{eq:exp_ens_ener}
	\E{}{\bw} = \sum_{I=0}^{\nEns-1} \ew{I} \E{}{(I)}
\end{equation}
can be obtained from the GOK variational principle
as follows\cite{Gross_1988a}
\begin{eqnarray}\label{eq:ens_energy}
	\E{}{\bw}  = \min_{\hGam{\bw}} \Tr[\hGam{\bw} \hH],
\end{eqnarray}
where $\hH = \hT + \hWee + \hVne$ contains the kinetic,
electron-electron and nuclei-electron interaction potential operators,
respectively, $\Tr$ denotes the trace, and $\hGam{\bw}$ is a trial
density matrix operator of the form
\begin{eqnarray}
	\hGam{\bw} = \sum_{I=0}^{\nEns - 1} \ew{I} \dyad*{\overline{\Psi}^{(I)}},
\end{eqnarray}
where $\lbrace \overline{\Psi}^{(I)} \rbrace_{0 \le I \le \nEns-1}$ is a set of $\nEns$ orthonormal trial wave functions. 
The lower bound of Eq.~\eqref{eq:ens_energy} is reached when the set of wave functions correspond to the exact eigenstates of $\hH$, \ie, $\lbrace \overline{\Psi}^{(I)} \rbrace_{0 \le I \le \nEns-1} = \lbrace \Psi^{(I)} \rbrace_{0 \le I \le \nEns-1}$.
Multiplet degeneracies can be easily handled by assigning the same
weight to the degenerate states. \cite{Gross_1988b}
One of the key feature of the GOK ensemble is that excitation
energies can be extracted from the ensemble energy via differentiation
with respect to the individual excited-state weights $\ew{I}$ ($I>0$):
\begin{equation}\label{eq:diff_Ew}
	\pdv{\E{}{\bw}}{\ew{I}} = \E{}{(I)} - \E{}{(0)} = \Ex{}{(I)}.
\end{equation}

Turning to GOK-DFT, the extension of the Hohenberg--Kohn theorem to ensembles allows to rewrite the exact variational expression for the ensemble energy as\cite{Gross_1988b}
\begin{equation}
\label{eq:Ew-GOK}
	\E{}{\bw} = \min_{\n{}{}} \qty{ \F{}{\bw}[\n{}{}] + \int \vne(\br{}) \n{}{}(\br{}) d\br{} },
\end{equation}
where $\vne(\br{})$ is the external potential and $\F{}{\bw}[\n{}{}]$ is the universal ensemble functional
(the weight-dependent analog of the Hohenberg--Kohn universal functional for ensembles).
In the KS formulation\cite{Gross_1988b}, this functional can be decomposed as
\begin{equation}\label{eq:FGOK_decomp}
      \F{}{\bw}[\n{}{}] 
      = \Tr{ \hgamdens{\bw} \hT }+ \E{\Ha}{}[\n{}{}]+\E{\xc}{\bw}[\n{}{}],
\end{equation}
where
$\Tr{ \hgamdens{\bw} \hT } =\Ts{\bw}[\n{}{}]$ is the noninteracting ensemble kinetic energy functional,
\begin{equation}
	\hgam{\bw}[n] = \sum_{I=0}^{\nEns-1} \ew{I} \dyad{\Det{I}{\bw}[n]}
\end{equation}
is the density-functional KS density matrix operator, and $\lbrace
\Det{I}{\bw}[n] \rbrace_{0 \le I \le \nEns-1}$ are single-determinant
wave functions (or configuration state functions\cite{Gould_2017}). 
Their dependence on the density is determined from the ensemble density constraint 
\begin{equation}
	\sum_{I=0}^{\nEns-1} \ew{I} \n{\Det{I}{\bw}[n]}{}(\br) = \n{}{}(\br).
\end{equation}
Note that the original decomposition \cite{Gross_1988b} shown in Eq.~\eqref{eq:FGOK_decomp}, where the
conventional (weight-independent) Hartree functional
\beq \label{eq:Hartree}
	\E{\Ha}{}[\n{}{}]=\frac{1}{2} \iint \frac{\n{}{}(\br{}) \n{}{}(\br{}')}{\abs{\br{}-\br{}'}} d\br{} d\br{}'
\eeq
is separated
from the (weight-dependent) exchange-correlation (xc) functional, is
formally exact. In practice, the use of such a decomposition might be
problematic as inserting an ensemble density into $\E{\Ha}{}[\n{}{}]$
causes the infamous ghost-interaction error. \cite{Gidopoulos_2002,Pastorczak_2014, Alam_2016, Alam_2017, Gould_2017} 
The latter should in principle be removed by the exchange component of the ensemble xc functional 
$\E{\xc}{\bw}[\n{}{}] \equiv \E{\ex}{\bw}[\n{}{}] + \E{\co}{\bw}[\n{}{}]$, 
as readily seen from the exact expression 
\beq
	\E{\ex}{\bw}[\n{}{}] 
	= \sum_{I=0}^{\nEns-1} \ew{I}\mel{\Det{I}{\bw}[\n{}{}]}{\hWee}{\Det{I}{\bw}[\n{}{}]} - \E{\Ha}{}[\n{}{}].
\eeq
The minimum in Eq.~\eqref{eq:Ew-GOK} is reached when the density $n$
equals the exact ensemble one 
\beq\label{eq:nw}
n^{\bw}(\br)=\sum_{I=0}^{\nEns-1}
\ew{I}n_{\Psi_I}(\br).
\eeq
In practice, the minimising KS density matrix operator
$\hgam{\bw}\left[\n{}{\bw}\right]$ 
can be determined from the following KS reformulation of the
GOK variational principle, \cite{Gross_1988b,Senjean_2015} 
\beq\label{eq:min_KS_DM}
\E{}{\bw}  = \min_{\hGam{\bw}} \left\{\Tr[\hGam{\bw}
\left(\hT+\hVne\right)]+\E{\Ha}{}[\n{\hGam{\bw}}{}]+\E{\xc}{\bw}[\n{\hGam{\bw}}{}]\right\},
\eeq 
where $\n{\hGam{\bw}}{}(\br)=\sum_{I=0}^{\nEns - 1}
\ew{I}\n{\overline{\Psi}^{(I)}}{}$ is a trial ensemble density. As a
result, the orbitals
$\lbrace \MO{p}{\bw}(\br{}) \rbrace_{1 \le p \le
\nOrb}$ from which the KS
wave functions $\left\{\Det{I}{\bw}\left[n^{\bw}\right]\right\}_{0\leq
I\leq \nEns-1}$ are constructed can be obtained by solving the following ensemble KS equation
\begin{equation}
\label{eq:eKS}
	\qty{ \hHc(\br{}) + \fdv{\E{\Hxc}{\bw}[\n{}{\bw}]}{\n{}{}(\br{})}} \MO{p}{\bw}(\br{}) = \eps{p}{\bw} \MO{p}{\bw}(\br{}),
\end{equation}
where $\hHc(\br{}) = -\nabla^2/2  + \vne(\br{})$, and
\begin{equation}
	\fdv{\E{\Hxc}{\bw}[\n{}{}]}{\n{}{}(\br{})} 
	= 
\int \frac{\n{}{}(\br{}')}{\abs{\br{}-\br{}'}} d\br{}'
+ \fdv{\E{\xc}{\bw}[\n{}{}]}{\n{}{}(\br{})}.
\end{equation}
The ensemble density can be obtained directly (and exactly, if no
approximation is made) from these orbitals, \ie,
\beq\label{eq:ens_KS_dens}
\n{}{\bw}(\br{})=\sum_{I=0}^{\nEns-1} \ew{I}\left(\sum_{p}^{\nOrb}
\ON{p}{(I)} [\MO{p}{\bw}(\br{})]^2\right), 
\eeq
where $\ON{p}{(I)}$ denotes the occupation of $\MO{p}{\bw}(\br{})$ in
the $I$th KS wave function $\Det{I}{\bw}\left[n^{\bw}\right]$. Turning
to the excitation energies, they can be extracted from the
density-functional ensemble as follows [see Eqs.~\eqref{eq:diff_Ew}
and \eqref{eq:min_KS_DM} and Refs.~\onlinecite{Gross_1988b,Deur_2019}]:
\beq
	\label{eq:dEdw}
	\Omega^{(I)}= \Eps{I}{\bw} - \Eps{0}{\bw} + \left. \pdv{\E{\xc}{\bw}[\n{}{}]}{\ew{I}} \right|_{\n{}{} = \n{}{\bw}},
\eeq
where
\begin{equation}
\label{eq:KS-energy}
	\Eps{I}{\bw} = \sum_{p}^{\nOrb} \ON{p}{(I)} \eps{p}{\bw}
\end{equation}
is the energy of the $I$th KS state.

Equation \eqref{eq:dEdw} is our working equation for computing excitation energies from a practical point of view.
Note that the individual KS densities
$\n{\Det{I}{\bw}\left[n^{\bw}\right]}{}(\br{})=\sum_{p}^{\nOrb}
\ON{p}{(I)} [\MO{p}{\bw}(\br{})]^2$ do
not necessarily match the \textit{exact} (interacting) individual-state
densities $n_{\Psi_I}(\br)$ as the non-interacting KS ensemble is expected to reproduce the true interacting ensemble density $\n{}{\bw}(\br{})$ defined in Eq.~\eqref{eq:nw}, and not each individual density. 
Nevertheless, these densities can still be extracted in principle
exactly from the KS ensemble as shown by one of the author. \cite{Fromager_2020}

In the following, we will work at the (weight-dependent) ensemble LDA (eLDA) level of approximation, \ie
\beq
\E{\xc}{\bw}[\n{}{}] 
&\overset{\rm eLDA}{\approx}&
\int \e{\xc}{\bw}(\n{}{}(\br{})) \n{}{}(\br{}) d\br{},
\\
 \fdv{\E{\xc}{\bw}[\n{}{}]}{\n{}{}(\br{})} 
&\overset{\rm eLDA}{\approx}&
\left. \pdv{\e{\xc}{\bw{}}(\n{}{})}{\n{}{}} \right|_{\n{}{} = \n{}{}(\br{})} \n{}{}(\br{}) + \e{\xc}{\bw{}}(\n{}{}(\br{})).
\eeq
We will also adopt the usual decomposition, and write down the weight-dependent xc functional as
\begin{equation}
	\e{\xc}{\bw{}}(\n{}{}) = \e{\ex}{\bw{}}(\n{}{}) + \e{\co}{\bw{}}(\n{}{}),
\end{equation}
where $\e{\ex}{\bw{}}(\n{}{})$ and $\e{\co}{\bw{}}(\n{}{})$ are the
weight-dependent density-functional exchange and correlation energies
per particle, respectively.
As shown in Sec.~\ref{subsubsec:weight-dep_corr_func}, the weight
dependence of the correlation energy can be extracted from a FUEG model. In order to make the resulting weight-dependent
correlation functional truly universal, \ie,
independent on the number of electrons in the FUEG, one could use the
curvature of the Fermi hole~\cite{Loos_2017a} as an additional variable in the
density-functional approximation. The development of such a 
generalised correlation eLDA is left for future work. Even though a similar strategy could be applied
to the weight-dependent exchange part, we
explore in the present work a different path where the
(system-dependent) exchange functional
parameterisation relies on the ensemble energy linearity
constraint (see Sec.~\ref{subsubsec:weight-dep_x_fun}). Finally, let us
stress that, in order to further
improve the description of the ensemble correlation energy, a
post-treatment of the recently
revealed density-driven
correlations~\cite{Gould_2019a,Gould_2019b,Gould_2020,Fromager_2020} (which, by construction, are absent
from FUEGs) might be necessary. An orbital-dependent correction derived
in Ref.~\onlinecite{Fromager_2020} might be
used for that purpose. Work is currently in progress in this 
direction.

\section{Computational details}
\label{sec:compdet}

The self-consistent GOK-DFT calculations [see Eqs.~\eqref{eq:eKS} and \eqref{eq:ens_KS_dens}] have been performed in a restricted formalism with the \texttt{QuAcK} software, \cite{QuAcK} freely available on \texttt{github}, where the present weight-dependent functionals have been implemented. 
For more details about the self-consistent implementation of GOK-DFT, we refer the interested reader to Ref.~\onlinecite{Loos_2020} where additional technical details can be found.
For all calculations, we use the aug-cc-pVXZ (X = D, T, Q, and 5) Dunning family of atomic basis sets. \cite{Dunning_1989,Kendall_1992,Woon_1994}
Numerical quadratures are performed with the \texttt{numgrid} library \cite{numgrid} using 194 angular points (Lebedev grid) and a radial precision of $10^{-7}$. \cite{Becke_1988b,Lindh_2001}

This study deals only with spin-unpolarised systems, \ie, $\n{\uparrow}{} = \n{\downarrow}{} = \n{}{}/2$ (where $\n{\uparrow}{}$ and $\n{\downarrow}{}$ are the spin-up and spin-down electron densities).
Moreover, we restrict our study to the case of a three-state ensemble (\ie, $\nEns = 3$) where the ground state ($I=0$ with weight $1 - \ew{1} - \ew{2}$), a singly-excited state ($I=1$ with weight $\ew{1}$), as well as the lowest doubly-excited state ($I=2$ with weight $\ew{2}$) are considered.
Assuming that the singly-excited state is lower in energy than the doubly-excited state, one should have $0 \le \ew{2} \le 1/3$ and $\ew{2} \le \ew{1} \le (1 - \ew{2})/2$ to ensure the GOK variational principle.
If the doubly-excited state (whose weight is denoted $\ew{2}$
throughout this work) is lower in energy than the singly-excited state
(with weight $\ew{1}$), which can be the case as one would notice
later, then one has to swap $\ew{1}$ and $\ew{2}$ in the above
inequalities. 
Note also that additional lower-in-energy single excitations may have to be included into the ensemble before incorporating the double excitation of interest. 
In the present exploratory work, we will simply exclude them from the ensemble and leave the more consistent (from a GOK point of view) description of all low-lying excitations to future work.
Unless otherwise stated, we set the same weight to the two excited states (\ie, $\ew{} \equiv \ew{1} = \ew{2}$).
In this case, the ensemble energy will be written as a single-weight quantity, $\E{}{\ew{}}$.
The zero-weight limit (\ie, $\ew{} \equiv \ew{1} = \ew{2} = 0$), and the equi-weight ensemble (\ie, $\ew{} \equiv \ew{1} = \ew{2} = 1/3$) are considered in the following.
(Note that the zero-weight limit corresponds to a conventional ground-state KS calculation.)

Let us finally mention that we will sometimes ``violate'' the GOK
variational principle in order to build our weight-dependent functionals
by considering the extended range of weights $0 \le \ew{2} \le 1$.
The pure-state limit, $\ew{1} = 0 \land \ew{2} = 1$, is of particular interest as it is, like the
(ground-state) zero-weight limit, a genuine saddle point of the
restricted KS equations [see Eqs.~\eqref{eq:min_KS_DM} and
\eqref{eq:eKS}], and it matches perfectly the results obtained with the maximum overlap method (MOM) developed by Gilbert, Gill and coworkers. \cite{Gilbert_2008,Barca_2018a,Barca_2018b}
From a GOK-DFT perspective, considering a (stationary) pure-excited-state limit can be seen as a way to construct density-functional approximations to individual exchange and state-driven correlation within an ensemble. \cite{Gould_2019a,Gould_2019b,Fromager_2020}
However, when it comes to compute excitation energies, we will exclusively consider ensembles where the largest weight is assigned to the ground state.

\section{Results and Discussion}
\label{sec:res}
In this Section, we propose a two-step procedure to design, first, a weight- and system-dependent local exchange functional in order to remove some of the curvature of the ensemble energy.
Second, we describe the construction of a universal, weight-dependent local correlation functional based on FUEGs.
This procedure is applied to various two-electron systems in order to extract excitation energies associated with doubly-excited states.

\subsection{Hydrogen molecule at equilibrium}
\label{sec:H2}

\subsubsection{Weight-independent exchange functional}

First, we compute the ensemble energy of the \ce{H2} molecule at
equilibrium bond length (\ie, $\RHH = 1.4$ bohr) using the aug-cc-pVTZ
basis set and the conventional (weight-independent) LDA Slater exchange functional (\ie, no correlation functional is employed), \cite{Dirac_1930, Slater_1951} which is explicitly given by 
\begin{align}
\label{eq:Slater}
	\e{\ex}{\text{S}}(\n{}{}) & = \Cx{} \n{}{1/3},
	&
	\Cx{} & = -\frac{3}{4} \qty(\frac{3}{\pi})^{1/3}.
\end{align}
In the case of \ce{H2}, the ensemble is composed by the
ground state of electronic configuration $1\sigma_g^2$, the lowest
singly-excited state of
configuration $1\sigma_g 2\sigma_g$, and the lowest doubly-excited state
of configuration $1\sigma_u^2$ (which has an auto-ionising resonance nature \cite{Bottcher_1974})
which all are of symmetry $\Sigma_g^+$.
As mentioned previously, the lower-lying
singly-excited states like $1\sigma_g3\sigma_g$ and
$1\sigma_g4\sigma_g$, which should in principle be part of the ensemble
(see Fig.~3 in Ref.~\onlinecite{TDDFTfromager2013}),
have been excluded, for simplicity.

\begin{figure}
	\includegraphics[width=\linewidth]{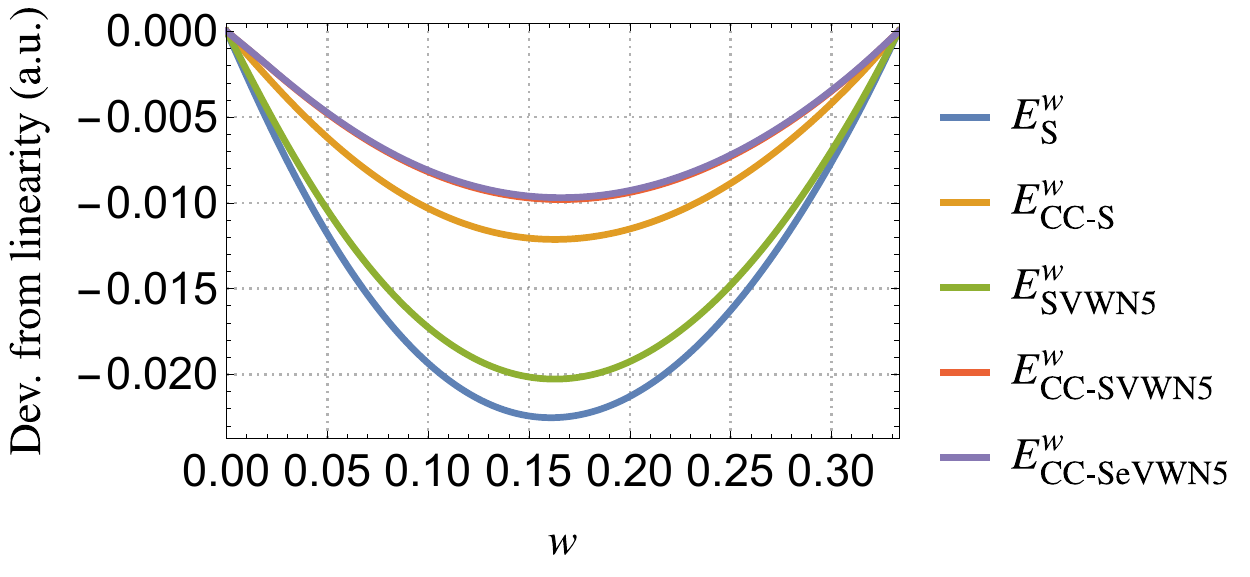}
	\caption{
	\ce{H2} at equilibrium bond length: deviation from linearity of the ensemble energy $\E{}{\ew{}}$ (in hartree) as a function of $\ew{}$ for various functionals and the aug-cc-pVTZ basis set.
	See main text for the definition of the various functionals' acronyms.
	\label{fig:Ew_H2}
	}
\end{figure}

\begin{figure}
	\includegraphics[width=\linewidth]{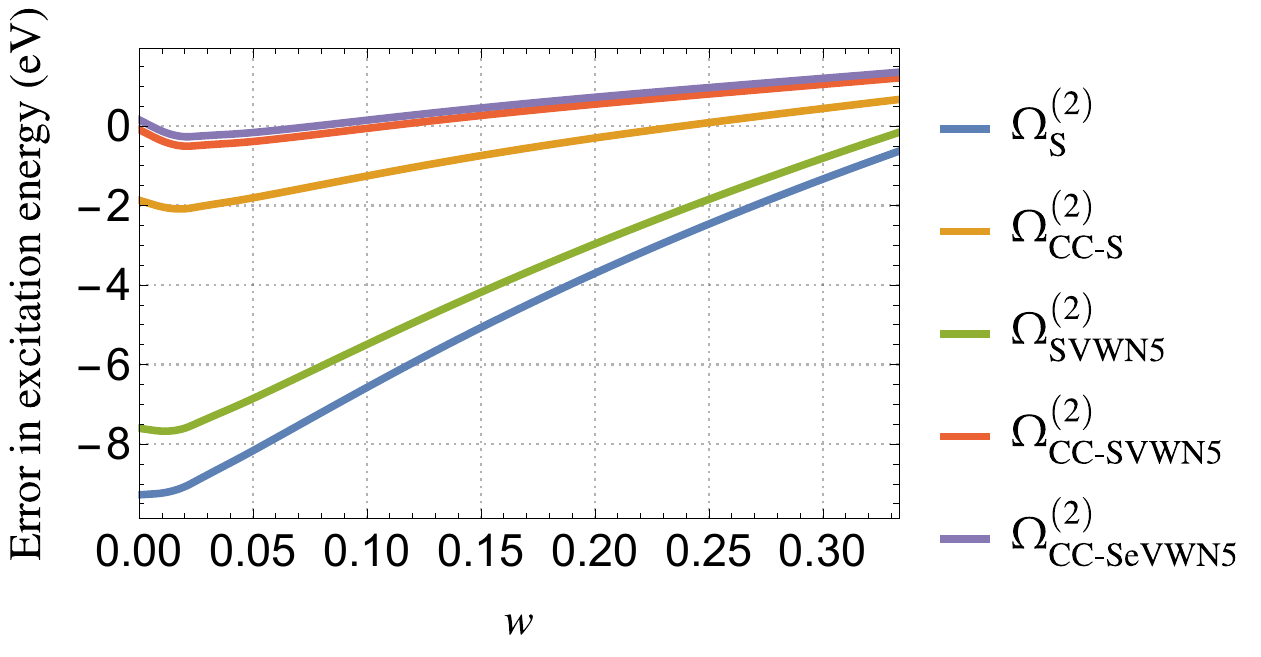}
	\caption{
	\ce{H2} at equilibrium bond length: error (with respect to FCI) in the excitation energy $\Ex{}{(2)}$ (in eV) associated with the doubly-excited state as a function of $\ew{}$ for various functionals and the aug-cc-pVTZ basis set.
	See main text for the definition of the various functionals' acronyms.
	\label{fig:Om_H2}
	}
\end{figure}

The deviation from linearity of the ensemble energy $\E{}{\ew{}}$
[we recall that $\ew{1}=\ew{2}=\ew{}$] is depicted in Fig.~\ref{fig:Ew_H2} as a function of weight $0 \le \ew{} \le 1/3$ (blue curve).
Because the Slater exchange functional defined in Eq.~\eqref{eq:Slater} does not depend on the ensemble weight, there is no contribution from the ensemble derivative term [last term in Eq.~\eqref{eq:dEdw}].
As anticipated, $\E{}{\ew{}}$ is far from being linear, which means that
the excitation energy associated with the doubly-excited state obtained
via the derivative of the ensemble energy with respect to $\ew{2}$
(and taken at $\ew{2}=\ew{}=\ew{1}$) varies significantly with $\ew{}$ (see blue curve in Fig.~\ref{fig:Om_H2}).
Taking as a reference the full configuration interaction (FCI) value of $28.75$ eV obtained with the aug-mcc-pV8Z basis set, \cite{Barca_2018a} one can see that the excitation energy varies by more than $8$ eV from $\ew{} = 0$ to $1/3$.
Note that the exact xc ensemble functional would yield a perfectly
linear ensemble energy and, hence, the same value of the excitation energy independently of the ensemble weights.

\subsubsection{Weight-dependent exchange
functional}\label{subsubsec:weight-dep_x_fun}

Second, in order to remove some of this spurious curvature of the ensemble
energy (which is mostly due to the ghost-interaction error, \cite{Gidopoulos_2002} but not only \cite{Loos_2020}), 
one can easily reverse-engineer (for this particular system, geometry, basis set, and excitation) a local exchange functional to make $\E{}{(0,\ew{2})}$ as linear as possible for $0 \le \ew{2} \le 1$ assuming a perfect linearity between the pure-state limits $ \ew{1} = \ew{2} = 0$ (ground state) and $\ew{1} = 0 \land \ew{2} = 1$ (doubly-excited state).
Doing so, we have found that the following weight-dependent exchange functional (denoted as CC-S for ``curvature-corrected'' Slater functional)
\begin{equation}\label{eq:ensemble_Slater_func}
	\e{\ex}{\ew{2},\text{CC-S}}(\n{}{}) = \Cx{\ew{2}} \n{}{1/3},
\end{equation}
with 
\begin{equation}
\label{eq:Cxw}
	\frac{\Cx{\ew{2}}}{\Cx{}} = 1 - \ew{2} (1 - \ew{2})\qty[ \alpha + \beta (\ew{2} - 1/2) + \gamma (\ew{2} - 1/2)^2 ],
\end{equation}
and 
\begin{subequations}
\begin{align}
	\alpha & = + 0.575\,178,
	&
	\beta & = - 0.021\,108,
	&
	\gamma & = - 0.367\,189,
\end{align}
\end{subequations}
makes the ensemble energy $\E{}{(0,\ew{2})}$ almost perfectly linear (by construction), and removes some of the curvature of $\E{}{\ew{}}$ (see yellow curve in Fig.~\ref{fig:Ew_H2}).
It also allows to ``flatten the curve'' making the excitation energy much more stable (with respect to
$\ew{}$), and closer to the FCI reference (see yellow curve in
Fig.~\ref{fig:Om_H2}).\\

The parameters $\alpha$, $\beta$, and $\gamma$ entering Eq.~\eqref{eq:Cxw} have been obtained via a least-square fit of the non-linear component of the ensemble energy computed between $\ew{2} = 0$ and $\ew{2} = 1$ by steps of $0.025$. 
Although this range of weights is inconsistent with GOK theory, we have found that it is important, from a practical point of view, to ensure a correct behaviour in the whole range of weights in order to obtain accurate excitation energies.
Note that the CC-S functional depends on $\ew{2}$ only, and not $\ew{1}$, as it is specifically tuned for the double excitation.
Hence, only the double excitation includes a contribution from the ensemble derivative term [see Eq.~\eqref{eq:dEdw}].

\begin{figure}
	\includegraphics[width=\linewidth]{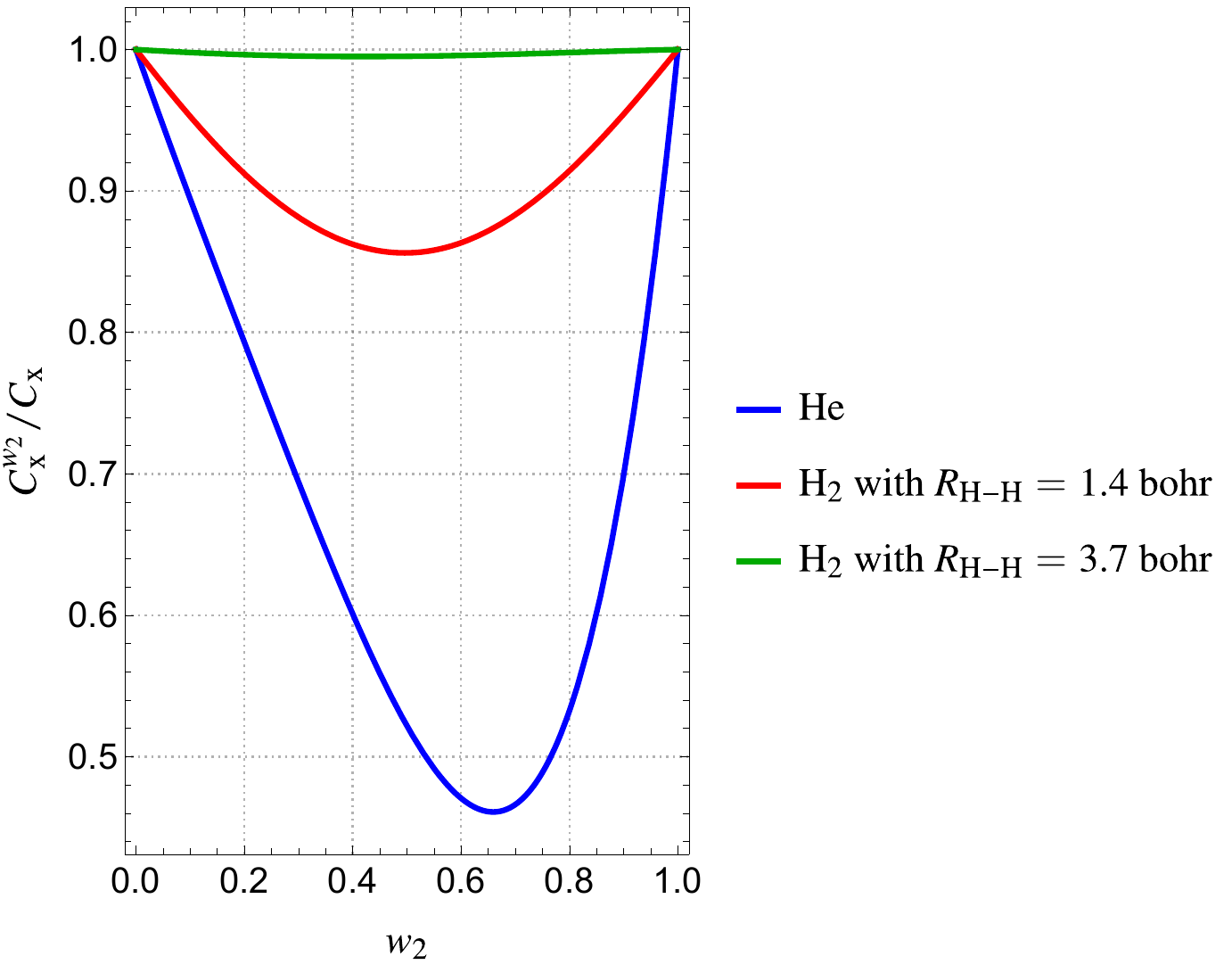}
	\caption{
	$\Cx{\ew{2}}/\Cx{}$ as a function of $\ew{2}$ [see Eq.~\eqref{eq:Cxw}] computed with the aug-cc-pVTZ basis set for the \ce{He} atom (blue) and the \ce{H2} molecule at $\RHH = 1.4$ bohr (red), and $\RHH = 3.7$ bohr (green).
	\label{fig:Cxw}
	}
\end{figure}

The present procedure can be related to optimally-tuned range-separated hybrid functionals, \cite{Stein_2009} where the range-separation parameters (which control the amount of short- and long-range exact exchange) are determined individually for each system by iteratively tuning them in order to enforce non-empirical conditions related to frontier orbitals (\eg, ionisation potential, electron affinity, etc) or, more importantly here, the piecewise linearity of the ensemble energy for ensemble states described by a fractional number of electrons. \cite{Stein_2009,Stein_2010,Stein_2012,Refaely-Abramson_2012}
In this context, the analog of the ``ionisation potential theorem'' for the first
(neutral) 
excitation, for example, would read as follows [see
Eqs.~\eqref{eq:exp_ens_ener}, \eqref{eq:diff_Ew}, and \eqref{eq:dEdw}]:
\beq
2\left(E^{\ew{1}=1/2}-E^{\ew{1}=0}\right)&\overset{0\leq \ew{1}\leq 1/2}{=}&\Eps{1}{\ew{1}} - \Eps{0}{\ew{1}} + \left.
\pdv{\E{\xc}{\ew{1}}[\n{}{}]}{\ew{1}} \right|_{\n{}{} =
\n{}{\ew{1}}}.
\eeq
We enforce this type of \textit{exact} constraint (to the maximum possible extent) when optimising the parameters in Eq.~\eqref{eq:Cxw} in order to minimise the curvature of the ensemble energy.
As readily seen from Eq.~\eqref{eq:Cxw} and graphically illustrated in Fig.~\ref{fig:Cxw} (red curve), the weight-dependent correction does not affect the two ghost-interaction-free limits at $\ew{1} = \ew{2} = 0$ and $\ew{1} = 0 \land \ew{2} = 1$ (\ie, the pure-state limits), as $\Cx{\ew{2}}$ reduces to $\Cx{}$ in these two limits.
Indeed, it is important to ensure that the weight-dependent functional does not alter these pure-state limits, which are genuine saddle points of the KS equations, as mentioned above.
Finally, let us mention that, around $\ew{2} = 0$, the behaviour of Eq.~\eqref{eq:Cxw} is linear: this is the main feature that one needs to catch in order to get accurate excitation energies in the zero-weight limit which is ghost-interaction free.
Nonetheless, beyond the $\ew{2} = 0$ limit, the CC-S functional also includes quadratic terms in order to compensate the spurious curvature of the ensemble energy originating, mainly, from the Hartree term [see Eq.~\eqref{eq:Hartree}].

\subsubsection{Weight-independent correlation functional}

Third, we include correlation effects via the conventional VWN5 local correlation functional. \cite{Vosko_1980}
For the sake of clarity, the explicit expression of the VWN5 functional is not reported here but it can be found in Ref.~\onlinecite{Vosko_1980}.
The combination of the (weight-independent) Slater and VWN5 functionals (SVWN5) yield a highly convex ensemble energy (green curve in Fig.~\ref{fig:Ew_H2}), while the combination of CC-S and VWN5 (CC-SVWN5) exhibit a smaller curvature and improved excitation energies (red curve in Figs.~\ref{fig:Ew_H2} and \ref{fig:Om_H2}), especially at small weights, where the CC-SVWN5 excitation energy is almost spot on. 

\subsubsection{Weight-dependent correlation
functional}\label{subsubsec:weight-dep_corr_func}

Fourth, in the spirit of our recent work, \cite{Loos_2020} we design a universal, weight-dependent correlation functional. 
To build this correlation functional, we consider the singlet ground state, the first singly-excited state, as well as the first doubly-excited state of a two-electron FUEGs which consists of two electrons confined to the surface of a 3-sphere (also known as a glome). \cite{Loos_2009a,Loos_2009c,Loos_2010e}
Notably, these three states have the same (uniform) density $\n{}{} = 2/(2\pi^2 R^3)$, where $R$ is the radius of the 3-sphere onto which the electrons are confined.
Note that the present paradigm is equivalent to the conventional IUEG model in the thermodynamic limit. \cite{Loos_2011b}
We refer the interested reader to Refs.~\onlinecite{Loos_2011b,Loos_2017a} for more details about this paradigm.

The reduced (\ie, per electron) Hartree-Fock (HF) energies for these three states are
\begin{subequations}
\begin{align}
	\e{\HF}{(0)}(\n{}{}) & = \frac{4}{3} \qty(\frac{\n{}{}}{\pi})^{1/3},
	\label{eq:eHF_0}
	\\
	\e{\HF}{(1)}(\n{}{}) & = \frac{3\pi^{2}}{4} \qty(\frac{\n{}{}}{\pi})^{2/3} + \frac{16}{10} \qty(\frac{\n{}{}}{\pi})^{1/3}.
	\label{eq:eHF_1}
	\\
	\e{\HF}{(2)}(\n{}{}) & = \frac{3\pi^{2}}{2} \qty(\frac{\n{}{}}{\pi})^{2/3} + \frac{176}{105} \qty(\frac{\n{}{}}{\pi})^{1/3}.
	\label{eq:eHF_2}
\end{align}
\end{subequations}

\begin{figure}
	\includegraphics[width=0.8\linewidth]{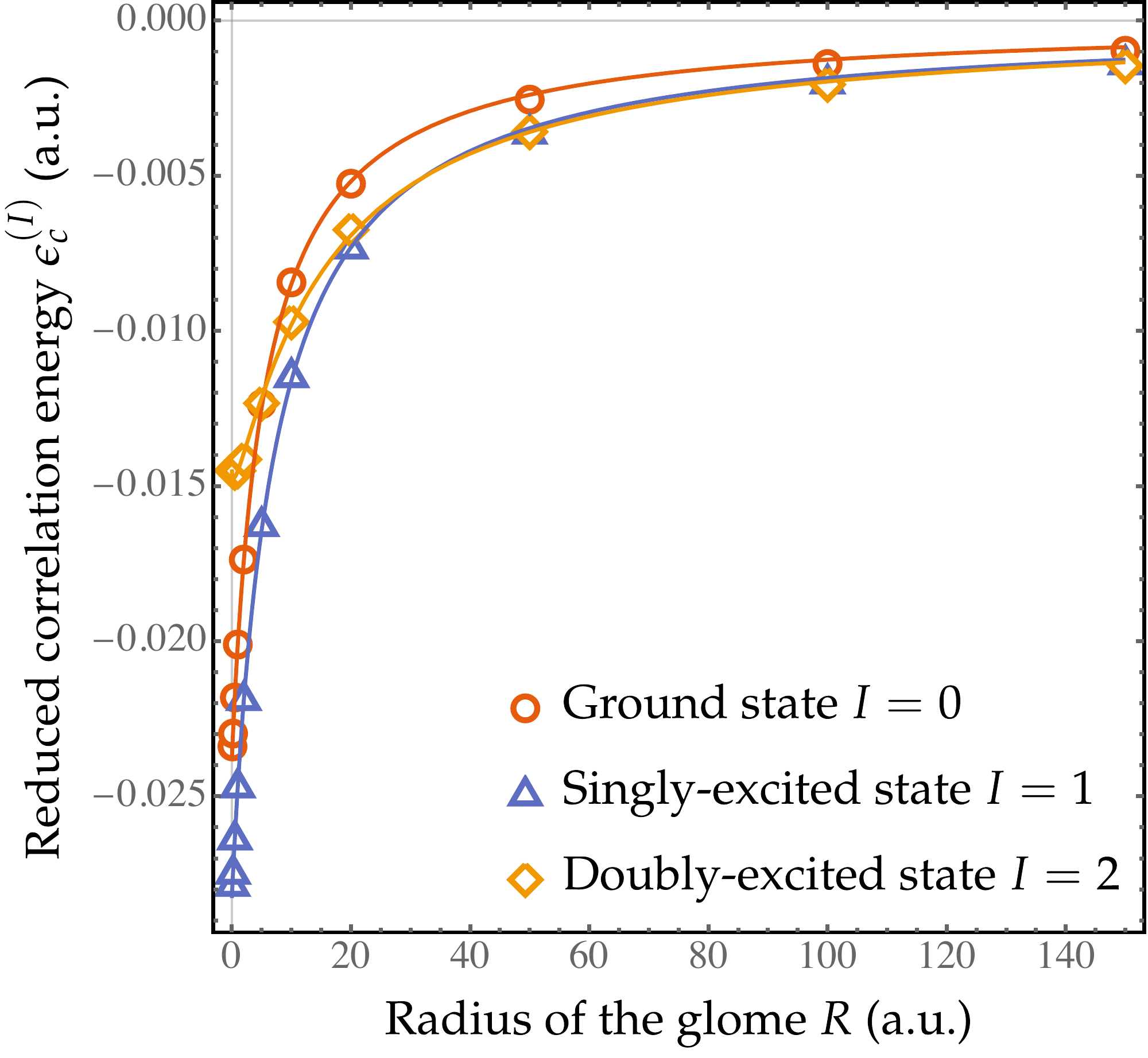}
	\caption{
		Reduced (i.e., per electron) correlation energy $\e{\co}{(I)}$ [see Eq.~\eqref{eq:ec}] as a function of $R = 1/(\pi^2 \n{}{})^{1/3}$ for the ground state ($I=0$), the first singly-excited state ($I=1$), and the first doubly-excited state ($I=2$) of the two-electron FUEG.
	The data gathered in Table \ref{tab:Ref} are also reported. 
	}
	\label{fig:Ec}
\end{figure}

Thanks to highly-accurate calculations \cite{Loos_2009a,Loos_2009c,Loos_2010e} and the expressions of the HF energies provided by Eqs.~\eqref{eq:eHF_0}, \eqref{eq:eHF_1}, and \eqref{eq:eHF_2}, one can write down, for each state, an accurate analytical expression of the reduced correlation energy \cite{Loos_2013a, Loos_2014a} via the following simple Pad\'e approximant \cite{Sun_2016,Loos_2020}
\begin{equation}
\label{eq:ec}
	\e{\co}{(I)}(\n{}{}) = \frac{a_1^{(I)}}{1 + a_2^{(I)} \n{}{-1/6} + a_3^{(I)} \n{}{-1/3}},
\end{equation}
where $a_2^{(I)}$ and $a_3^{(I)}$ are state-specific fitting parameters, which are provided in Table \ref{tab:OG_func}.
The value of $a_1^{(I)}$ is obtained via the exact high-density expansion of the correlation energy. \cite{Loos_2011b}
Equation \eqref{eq:ec} is depicted in Fig.~\ref{fig:Ec} for each state alongside the data gathered in Table \ref{tab:Ref}.
Combining these, we build a three-state weight-dependent correlation functional:
\begin{equation}
\label{eq:ecw}
	\e{\co}{\bw}(\n{}{}) =  (1-\ew{1}-\ew{2}) \e{\co}{(0)}(\n{}{}) + \ew{1} \e{\co}{(1)}(\n{}{}) + \ew{2} \e{\co}{(2)}(\n{}{}),
\end{equation}
where, unlike in the exact theory, \cite{Fromager_2020} the individual components are weight \textit{independent}.

\begin{table}
	\caption{
	\label{tab:Ref}
	$-\e{\co}{(I)}$ as a function of the radius of the glome $R = 1/(\pi^2 \n{}{})^{1/3}$ for the ground state ($I=0$), the first singly-excited state ($I=1$), and the first doubly-excited state ($I=2$) of the two-electron FUEG.
	}
	\begin{ruledtabular}
	\begin{tabular}{lccc}
					&	\tabc{Ground state}	&	\tabc{Single excitation}	&	\tabc{Double excitation}	\\
			$R$		&	\tabc{$I=0$}		&	\tabc{$I=1$}				&	\tabc{$I=2$}	\\
			\hline
			$0$		&	$0.023\,818$		&	0.028\,281					&	$0.014\,463$	\\
			$0.1$	&	$0.023\,392$		&	0.027\,886					&	$0.014\,497$	\\
			$0.2$	&	$0.022\,979$		&	0.027\,499					&	$0.014\,523$	\\
			$0.5$	&	$0.021\,817$		&	0.026\,394					&	$0.014\,561$	\\
			$1$		&	$0.020\,109$		&	0.024\,718					&	$0.014\,512$	\\
			$2$		&	$0.017\,371$		&	0.021\,901					&	$0.014\,142$	\\
			$5$		&	$0.012\,359$		&	0.016\,295					&	$0.012\,334$	\\
			$10$	&	$0.008\,436$		&	0.011\,494					&	$0.009\,716$	\\
			$20$	&	$0.005\,257$		&	0.007\,349					&	$0.006\,744$	\\
			$50$	&	$0.002\,546$		&	0.003\,643					&	$0.003\,584$	\\
			$100$	&	$0.001\,399$		&	0.002\,025					&	$0.002\,059$	\\
			$150$	&	$0.000\,972$		&	0.001\,414					&	$0.001\,458$	\\
	\end{tabular}
	\end{ruledtabular}
\end{table}

\begin{table}
	\caption{
	\label{tab:OG_func}
	Parameters of the correlation functionals for each individual state defined in Eq.~\eqref{eq:ec}.
	The values of $a_1$ are obtained to reproduce the exact high density correlation energy of each individual state, while $a_2$ and $a_3$ are fitted on the numerical values reported in Table \ref{tab:Ref}.}
	\begin{ruledtabular}
		\begin{tabular}{llll}
							&	\tabc{Ground state}	&	\tabc{Single excitation}	&	\tabc{Double excitation}	\\
							&	\tabc{$I=0$}		&	\tabc{$I=1$}				&	\tabc{$I=2$}				\\
			\hline
				$a_1$		&	$-0.023\,818\,4$	&	$-0.028\,281\,4$			&	$-0.014\,463\,3$	\\
				$a_2$		&	$+0.005\,409\,94$	&	$+0.002\,739\,25$			&	$-0.050\,602\,0$	\\	
				$a_3$		&	$+0.083\,076\,6$	&	$+0.066\,491\,4$			&	$+0.033\,141\,7$	\\
		\end{tabular}
	\end{ruledtabular}
\end{table}

Because our intent is to incorporate into standard functionals (which are ``universal'' in the sense that they do not depend on the number of electrons) information about excited states that will be extracted from finite systems (whose properties may depend on the number of electrons), we employ a simple ``embedding'' scheme where the two-electron FUEG (the impurity) is embedded in the IUEG (the bath).
As explained further in Ref.~\onlinecite{Loos_2020}, this embedding procedure can be theoretically justified by the generalised adiabatic connection formalism for ensembles originally derived by Franck and Fromager. \cite{Franck_2014} 
The weight-dependence of the correlation functional is then carried
exclusively by the impurity [\ie, the functional defined in
\eqref{eq:ecw}], while the remaining effects are produced by the bath
(\ie, the usual ground-state LDA correlation functional).
 
Consistently with such a strategy, Eq.~\eqref{eq:ecw} is ``centred'' on its corresponding weight-independent VWN5 LDA reference
\begin{equation}
\label{eq:becw}
	\e{\co}{\bw{},\eVWN}(\n{}{}) =  (1-\ew{1}-\ew{2}) \be{\co}{(0)}(\n{}{}) + \ew{1} \be{\co}{(1)}(\n{}{}) + \ew{2} \be{\co}{(2)}(\n{}{})
\end{equation}
via the following global, state-independent shift:
\begin{equation}
	\be{\co}{(I)}(\n{}{}) = \e{\co}{(I)}(\n{}{}) + \e{\co}{\VWN}(\n{}{}) - \e{\co}{(0)}(\n{}{}).
\end{equation}
In the following, we name this weight-dependent correlation functional ``eVWN5'' as it is a natural extension of the VWN5 local correlation functional for ensembles.
Also, Eq.~\eqref{eq:becw} can be recast as
\begin{equation}
\label{eq:eLDA}
\begin{split}
	\e{\co}{\bw{},\eVWN}(\n{}{}) 
	& =  \e{\co}{\VWN}(\n{}{}) 
	\\
	& + \ew{1} \qty[\e{\co}{(1)}(\n{}{}) - \e{\co}{(0)}(\n{}{})] 
	+ \ew{2} \qty[\e{\co}{(2)}(\n{}{}) - \e{\co}{(0)}(\n{}{})]
\end{split}
\end{equation}
which nicely highlights the centrality of VWN5 in the present weight-dependent density-functional approximation for ensembles.
In particular, $\e{\co}{(0,0),\eVWN}(\n{}{}) = \e{\co}{\VWN}(\n{}{})$.
We note also that, by construction, we have
\begin{equation}
\label{eq:dexcdw}
	\pdv{\e{\co}{\bw{},\eVWN}(\n{}{})}{\ew{I}}
	= \e{\co}{(I)}(n) - \e{\co}{(0)}(n),
\end{equation}
showing that the weight correction is purely linear in eVWN5 and entirely dependent on the FUEG model.
Contrary to the CC-S exchange functional which only depends on $\ew{2}$, the eVWN5 correlation functional depends on both weights.

As shown in Fig.~\ref{fig:Ew_H2}, the CC-SeVWN5 ensemble energy (as a function of $\ew{}$) is very slightly less
concave than its CC-SVWN5 counterpart and it also improves (not by much)
the excitation energy (see purple curve in Fig.~\ref{fig:Om_H2}).

For a more qualitative picture, Table \ref{tab:BigTab_H2} reports excitation energies for various methods and basis sets.
In particular, we report the excitation energies obtained with GOK-DFT
in the zero-weight limit (\ie, $\ew{} = 0$) and for equi-weights (\ie, $\ew{} = 1/3$).
These excitation energies are computed using
Eq.~\eqref{eq:dEdw}.

For comparison, we also report results obtained with the linear interpolation method (LIM). \cite{Senjean_2015,Senjean_2016} 
The latter simply consists in extracting the excitation energies (which are
weight-independent, by construction) from the equi-ensemble energies, as
follows:
\begin{subequations}
\begin{align}
	\Ex{\LIM}{(1)} & = 2 \qty[\E{}{\bw{}=(1/2,0)} - \E{}{\bw{}=(0,0)}], \label{eq:LIM1}
	\\
	\Ex{\LIM}{(2)} & = 3 \qty[\E{}{\bw{}=(1/3,1/3)} - \E{}{\bw{}=(1/2,0)}] + \frac{1}{2} \Ex{\LIM}{(1)}. \label{eq:LIM2}
\end{align}
\end{subequations}
For a general expression with multiple (and possibly degenerate) states, we refer the reader to Eq.~(106) of Ref.~\onlinecite{Senjean_2015}, where LIM is shown to interpolate linearly the ensemble energy between equi-ensembles.
Note that two calculations are needed to get the first LIM excitation energy, with an additional equi-ensemble calculation for each higher excitation energy.

Additionally, MOM excitation energies \cite{Gilbert_2008,Barca_2018a,Barca_2018b}
\begin{subequations}
\begin{align}
	\Ex{\MOM}{(1)} & = \E{}{\bw{}=(1,0)} - \E{}{\bw{}=(0,0)}, \label{eq:MOM1}
	\\
	\Ex{\MOM}{(2)} & = \E{}{\bw{}=(0,1)} - \E{}{\bw{}=(0,0)}, \label{eq:MOM2}
\end{align}
\end{subequations}
which also require three separate calculations at a different set of ensemble weights, have been computed for further comparisons.

As readily seen in Eqs.~\eqref{eq:LIM1} and \eqref{eq:LIM2}, LIM is a recursive strategy where the first excitation energy has to be determined
in order to compute the second one.
In the above equations, we
assumed that the singly-excited state (with weight $\ew{1}$) is lower
in energy than the doubly-excited state (with weight $\ew{2}$).
If the ordering changes (like in the case of the stretched \ce{H2} molecule, see below), one should substitute $\E{}{\bw{}=(0,1/2)}$
by $\E{}{\bw{}=(1/2,0)}$ in Eqs.~\eqref{eq:LIM1} and \eqref{eq:LIM2} which then correspond to the excitation energies of the
doubly-excited and singly-excited states, respectively.
The same holds for the MOM excitation energies in 
Eqs.~\eqref{eq:MOM1} and \eqref{eq:MOM2}.


The results gathered in Table \ref{tab:BigTab_H2} show that the GOK-DFT excitation energies obtained with the CC-SeVWN5 functional at zero weights are the most accurate with an improvement of $0.25$ eV as compared to CC-SVWN5, which is due to the ensemble derivative contribution of the eVWN5 functional.
The CC-SeVWN5 excitation energies at equi-weights (\ie, $\ew{} = 1/3$) are less satisfactory, but still remain in good agreement with FCI.
Interestingly, the CC-S functional leads to a substantial improvement of the LIM excitation energy, getting closer to the reference value when no correlation functional is used. When correlation functionals are added (\ie, VWN5 or eVWN5), LIM tends to overestimate
the excitation energy by about $1$ eV but still performs better than when no correction of the curvature is considered.
It is also important to mention that the CC-S functional does not alter the MOM excitation energy as the correction vanishes in this limit (\textit{vide supra}).
Finally, although we had to design a system-specific, weight-dependent exchange functional to reach such accuracy, we have not used any high-level reference data (such as FCI) to tune our functional, the only requirement being the linearity of the ensemble energy (obtained with LDA exchange) between the ghost-interaction-free pure-state limits.

\begin{table}
\caption{
Excitation energies (in eV) associated with the lowest double excitation of \ce{H2} with $\RHH = 1.4$ bohr for various methods, combinations of xc functionals, and basis sets.
\label{tab:BigTab_H2}
}
\begin{ruledtabular}
\begin{tabular}{llccccc}
	\mc{2}{c}{xc functional} 		&					&	\mc{2}{c}{GOK}	\\
	\cline{1-2}												\cline{4-5}
	\tabc{x}	&	\tabc{c}		&	Basis			&	 $\ew{} = 0$	&	$\ew{} = 1/3$	&	LIM\fnm[1]		&	MOM\fnm[1]		\\
	\hline
	HF			&					&	aug-cc-pVDZ		&	35.59			&	33.33			&		&	28.65	\\
				&					&	aug-cc-pVTZ		&	35.01			&	33.51			&		&	28.65	\\
				&					&	aug-cc-pVQZ		&	34.66			&	33.54			&		&	28.65	\\
	\\
	HF			&	VWN5			&	aug-cc-pVDZ		&	37.83			&	33.86			&		&	29.17	\\
				&					&	aug-cc-pVTZ		&	37.61			&	33.99			&		&	29.17	\\
				&					&	aug-cc-pVQZ		&	37.07			&	34.01			&		&	29.17	\\
	\\
	HF			&	eVWN5			&	aug-cc-pVDZ		&	38.09			&	34.00			&		&	29.34	\\
				&					&	aug-cc-pVTZ		&	37.61			&	34.13			&		&	29.34	\\
				&					&	aug-cc-pVQZ		&	37.32			&	34.14			&		&	29.34	\\
	\\
	S			&					&	aug-cc-pVDZ		&	19.44			&	28.00			&	25.09	&	26.60	\\
				&					&	aug-cc-pVTZ		&	19.47			&	28.11			&	25.20	&	26.67	\\
				&					&	aug-cc-pVQZ		&	19.41			&	28.13			&	25.22	&	26.67	\\
	\\
	S			&	VWN5			&	aug-cc-pVDZ		&	21.04			&	28.49			&	25.90	&	27.10	\\
				&					&	aug-cc-pVTZ		&	21.14			&	28.58			&	25.99	&	27.17	\\
				&					&	aug-cc-pVQZ		&	21.13			&	28.59			&	26.00	&	27.17	\\
	\\
	S			&	eVWN5			&	aug-cc-pVDZ		&	21.28			&	28.64			&	25.99	&	27.27	\\
				&					&	aug-cc-pVTZ		&	21.39			&	28.74			&	26.08	&	27.34	\\
				&					&	aug-cc-pVQZ		&	21.38			&	28.75			&	26.09	&	27.34	\\
	\\
	CC-S		&					&	aug-cc-pVDZ		&	26.83			&	29.29			&	28.83	&	26.60	\\
				&					&	aug-cc-pVTZ		&	26.88			&	29.41			&	28.96	&	26.67	\\
				&					&	aug-cc-pVQZ		&	26.82			&	29.43			&	28.97	&	26.67	\\
	\\
	CC-S		&	VWN5			&	aug-cc-pVDZ		&	28.54			&	29.85			&	29.73	&	27.10	\\
				&					&	aug-cc-pVTZ		&	28.66			&	29.96			&	29.83	&	27.17	\\
				&					&	aug-cc-pVQZ		&	28.64			&	29.97			&	29.84	&	27.17	\\
	\\
	CC-S		&	eVWN5			&	aug-cc-pVDZ		&	28.78			&	29.99			&	29.82	&	27.27	\\
				&					&	aug-cc-pVTZ		&	28.90			&	30.10			&	29.92	&	27.34	\\
				&					&	aug-cc-pVQZ		&	28.89			&	30.11			&	29.93	&	27.34	\\
	\hline
	B			&	LYP				&	aug-mcc-pV8Z	&					&					&		&		28.42	\\
	B3			&	LYP				&	aug-mcc-pV8Z	&					&					&		&		27.77	\\
	HF			&	LYP				&	aug-mcc-pV8Z	&					&					&		&		29.18	\\
	HF			&					&	aug-mcc-pV8Z	&					&					&		&		28.65	\\
	\hline
	\mc{6}{l}{Accurate\fnm[2]}														&		28.75	\\
\end{tabular}
\end{ruledtabular}
\fnt[1]{Equations \eqref{eq:LIM2} and \eqref{eq:MOM2} are used where the first weight corresponds to the singly-excited state.}
\fnt[2]{FCI/aug-mcc-pV8Z calculation from Ref.~\onlinecite{Barca_2018a}.}
\end{table}

\subsection{Hydrogen molecule at stretched geometry}
\label{sec:H2st}

To investigate the weight dependence of the xc functional in the strong correlation regime, we now consider the \ce{H2} molecule in a stretched geometry ($\RHH = 3.7$ bohr).
Note that, for this particular geometry, the doubly-excited state becomes the lowest excited state with the same symmetry as the ground state.
Although we could safely restrict ourselves to a bi-ensemble composed by the ground state and the doubly-excited state, we eschew doing this and we still consider the same tri-ensemble defined in Sec.~\ref{sec:H2}.
Nonetheless, one should just be careful when reading the equations reported above, as they correspond to the case where the singly-excited state is lower in energy than the doubly-excited state.
We then follow the same protocol as in Sec.~\ref{sec:H2}, and considering again the aug-cc-pVTZ basis set, we design a CC-S functional for this system at $\RHH = 3.7$ bohr.
It yields $\alpha = +0.019\,226$, $\beta = -0.017\,996$, and $\gamma = -0.022\,945$ [see Eq.~\eqref{eq:Cxw}].
The weight dependence of $\Cx{\ew{2}}$ is illustrated in
Fig.~\ref{fig:Cxw} (green curve).

One clearly sees that the correction brought by CC-S is much more gentle than at $\RHH = 1.4$ bohr, which means that the ensemble energy obtained with the LDA exchange functional is much more linear at $\RHH = 3.7$ bohr.
Note that this linearity at $\RHH = 3.7$ bohr was also observed using weight-independent xc functionals in Ref.~\onlinecite{Senjean_2015}.
Table \ref{tab:BigTab_H2st} reports, for the aug-cc-pVTZ basis set (which delivers basis set converged results), the same set of calculations as in Table \ref{tab:BigTab_H2}.
As a reference value, we computed a FCI/aug-cc-pV5Z excitation energy of $8.69$ eV, which compares well with previous studies. \cite{Senjean_2015}
For $\RHH = 3.7$ bohr, it is much harder to get an accurate estimate of the excitation energy, the closest match being reached with HF exchange and VWN5 correlation at equi-weights.
As expected from the linearity of the ensemble energy, the CC-S functional coupled or not with a correlation functional yield extremely stable excitation energies as a function of the weight, with only a few tenths of eV difference between the zero- and equi-weights limits.
As a direct consequence of this linearity, LIM and MOM
do not provide any noticeable improvement on the excitation
energy.
Nonetheless, the excitation energy is still off by $3$ eV.
The fundamental theoretical reason of such a poor agreement is not clear but it might be that, in this strongly correlated regime,
the weight-dependent correlation functional plays a significant
role not caught by our approximation.

For additional comparison, we provide the excitation energy calculated by short-range multiconfigurational DFT in Ref.~\onlinecite{Senjean_2015}, using the (weight-independent) srLDA functional \cite{Toulouse_2004} and setting the range-separation parameter to $\mu = 0.4$ bohr$^{-1}$. 
The excitation energy improves by $1$ eV compared to the weight-independent SVWN5 functional, thus showing that treating the long-range part of the electron-electron repulsion by wave function theory plays a significant role.

\begin{table}
\caption{
Excitation energies (in eV) associated with the lowest double excitation of \ce{H2} at $\RHH = 3.7$ bohr obtained with the aug-cc-pVTZ basis set for various methods and combinations of xc functionals.
\label{tab:BigTab_H2st}
}
\begin{ruledtabular}
\begin{tabular}{llcccc}
	\mc{2}{c}{xc functional} 		&	\mc{2}{c}{GOK}	\\
	\cline{1-2}							\cline{3-4}
	\tabc{x}	&	\tabc{c}		&	 $\ew{} = 0$	&	$\ew{} = 1/3$	&	LIM\fnm[1]		&	MOM\fnm[1]		\\
	\hline
	HF			&					&	19.09			&	8.82			&	12.92	&	6.52	\\
	HF			&	VWN5			&	19.40			&	8.81			&	13.02	&	6.49	\\
	HF			&	eVWN5			&	19.59			&	8.95			&	13.11	&	\fnm[2]	\\
	S			&					&	5.31			&	5.67			&	5.46	&	5.56	\\
	S			&	VWN5			&	5.34			&	5.64			&	5.46	&	5.52	\\
	S			&	eVWN5			&	5.53			&	5.79			&	5.56	&	5.72	\\
	CC-S		&					&	5.55			&	5.72			&	5.56	&	5.56	\\
	CC-S		&	VWN5			&	5.58			&	5.69			&	5.57	&	5.52	\\
	CC-S		&	eVWN5			&	5.77			&	5.84			&	5.66	&	5.72	\\
	\hline
	B			&	LYP				&					&					&			&	5.28	\\
	B3			&	LYP				&					&					&			&	5.55	\\
	HF			&	LYP				&					&					&			&	6.68	\\
	\hline
	\mc{2}{l}{srLDA ($\mu = 0.4$) \fnm[3]} & 6.39		&					&	6.47	&	\\
	\hline
	\mc{5}{l}{Accurate\fnm[4]}															&	8.69	\\
\end{tabular}
\end{ruledtabular}
\fnt[1]{Equations \eqref{eq:LIM1} and \eqref{eq:MOM1} are used where the first weight corresponds to the doubly-excited state.}
\fnt[2]{KS calculation does not converge.}
\fnt[3]{Short-range multiconfigurational DFT/aug-cc-pVQZ calculations from Ref.~\onlinecite{Senjean_2015}.}
\fnt[4]{FCI/aug-cc-pV5Z calculation performed with QUANTUM PACKAGE. \cite{QP2}}
\end{table}

\subsection{Helium atom}
\label{sec:He}

As a final example, we consider the \ce{He} atom which can be seen as the limiting form of the \ce{H2} molecule for very short bond lengths.
Similar to \ce{H2}, our ensemble contains the ground state of configuration $1s^2$, the lowest singlet excited state of configuration $1s2s$, and the first doubly-excited state of configuration $2s^2$.
In \ce{He}, the lowest doubly-excited state is an auto-ionising resonance state, extremely high in energy and lies in the continuum. \cite{Madden_1963}
In Ref.~\onlinecite{Burges_1995}, highly-accurate calculations estimate an excitation energy of $2.126$ hartree for this $1s^2 \rightarrow 2s^2$ transition.
Nonetheless, it can be nicely described with a Gaussian basis set containing enough diffuse functions.
Consequently, we consider for this particular example the d-aug-cc-pVQZ basis set which contains two sets of diffuse functions.
The excitation energies associated with this double excitation computed with various methods and combinations of xc functionals are gathered in Table \ref{tab:BigTab_He}.

Before analysing the results, we would like to highlight the fact that there is a large number of singly-excited states lying in between the $1s2s$ and $2s^2$ states.
Therefore, the present ensemble is not consistent with GOK theory.
However, it is impossible, from a practical point of view, to take into account all these single excitations.
We then restrict ourselves to a tri-ensemble keeping in mind the possible theoretical loopholes of such a choice.

The parameters of the CC-S weight-dependent exchange functional (computed with the smaller aug-cc-pVTZ basis) are $\alpha = +1.912\,574$, $\beta = +2.715\,267$, and $\gamma = +2.163\,422$ [see Eq.~\eqref{eq:Cxw}], the curvature of the ensemble energy being more pronounced in \ce{He} than in \ce{H2} (blue curve in Fig.~\ref{fig:Cxw}).
The results reported in Table \ref{tab:BigTab_He} evidence this strong weight dependence of the excitation energies for HF or LDA exchange.

The CC-S exchange functional attenuates significantly this dependence, and when coupled with the eVWN5 weight-dependent correlation functional, the CC-SeVWN5 excitation energy at $\ew{} = 0$ is only $18$ millihartree off the reference value.
As in the case of \ce{H2}, the excitation energies obtained at
zero-weight are more accurate than at equi-weight, while the opposite
conclusion was made in Ref.~\onlinecite{Loos_2020}. 
This motivates further the importance of developing weight-dependent functionals that yields linear ensemble energies in order to get rid of the weight-dependency of the excitation energy. Here again, the LIM excitation energy using the CC-S functional is very accurate with only a 22 millihartree error compared to the reference value, while adding the correlation contribution to the functional tends to overestimate the excitation energy.
Hence, in the light of the results obtained in this paper, it seems that the weight-dependent curvature correction to the exchange functional has the largest impact on the accuracy of the excitation energies.

As a final comment, let us stress again that the present protocol does not rely on high-level calculations as the sole requirement for constructing the CC-S functional is the linearity of the ensemble energy with respect to the weight of the double excitation.

\begin{table}
\caption{
Excitation energies (in hartree) associated with the lowest double excitation of \ce{He} obtained with the d-aug-cc-pVQZ basis set for various methods and combinations of xc functionals.
\label{tab:BigTab_He}
}
\begin{ruledtabular}
\begin{tabular}{llcccc}
	\mc{2}{c}{xc functional} 			&	\mc{2}{c}{GOK}	\\
	\cline{1-2}												\cline{3-4}
	\tabc{x}	&	\tabc{c}			&	 $\ew{} = 0$	&	$\ew{} = 1/3$	&	LIM\fnm[1]		&	MOM\fnm[1]		\\
	\hline
	HF			&						&	1.874			&	2.212			&	2.123	&	2.142	\\
	HF			&	VWN5				&	1.988			&	2.260			&	2.190	&	2.193	\\
	HF			&	eVWN5				&	2.000			&	2.265			&	2.193	&	2.196	\\
	S			&						&	1.062			&	2.056			&	1.675	&	2.030	\\
	S			&	VWN5				&	1.163			&	2.104			&	1.735	&	2.079	\\
	S			&	eVWN5				&	1.174			&	2.109			&	1.738	&	2.083	\\
	CC-S		&						&	1.996			&	2.264			&	2.148	&	2.030	\\
	CC-S		&	VWN5				&	2.107			&	2.318			&	2.215	&	2.079	\\
	CC-S		&	eVWN5				&	2.108			&	2.323			&	2.218	&	2.083	\\
	\hline
	B			&	LYP					&					&					&			&	2.147	\\
	B3			&	LYP					&					&					&			&	2.150	\\
	HF			&	LYP					&					&					&			&	2.171	\\
	\hline
	\mc{2}{l}{Accurate\fnm[2]}			&					&					&			&	2.126	\\
\end{tabular}
\end{ruledtabular}
\fnt[1]{Equations \eqref{eq:LIM2} and \eqref{eq:MOM2} are used where the first weight corresponds to the singly-excited state.}
\fnt[1]{Explicitly-correlated calculations from Ref.~\onlinecite{Burges_1995}.}
\end{table}

\section{Conclusion}
\label{sec:ccl}
In the present article, we have discussed the construction of first-rung (\ie, local) weight-dependent exchange-correlation density-functional approximations for two-electron systems (\ce{He} and \ce{H2}) specifically designed for the computation of double excitations within GOK-DFT, a time-\textit{independent} formalism capable of extracting excitation energies via the derivative of the ensemble energy with respect to the weight of each excited state.

In the spirit of optimally-tuned range-separated hybrid functionals, we have found that the construction of a system- and excitation-specific weight-dependent local exchange functional can significantly reduce the curvature of the ensemble energy and improves excitation energies.
The present weight-dependent exchange functional, CC-S, specifically tailored for double excitations, only depends on the weight of the doubly-excited state, CC-S being independent on the weight of the singly-excited state.
We are currently investigating a generalisation of the present procedure in order to include a dependency on both weights in the exchange functional.

Although the weight-dependent correlation functional developed in this paper (eVWN5) performs systematically better than their weight-independent counterpart (VWN5), the improvement remains rather small.
To better understand the reasons behind this, it would be particularly interesting to investigate the influence of the self-consistent procedure, 
\ie, the variation in excitation energy when the \textit{exact} ensemble density (built with the exact individual densities) is used instead
of the self-consistent one. 
Exploring the impact of both density- and state-driven correlations \cite{Gould_2019a,Gould_2019b,Fromager_2020} may provide additional insights about the present results. 
This is left for future work.

In the light of the results obtained in this study on double excitations computed within the GOK-DFT framework, we believe that the development of more universal weight-dependent exchange and correlation functionals has a bright future, and we hope to be able to report further on this in the near future.

\begin{acknowledgements}
PFL thanks Radovan Bast and Anthony Scemama for technical assistance, as well as Julien Toulouse for stimulating discussions on double excitations.
CM thanks the \textit{Universit\'e Paul Sabatier} (Toulouse, France) for a PhD scholarship.
This work has also been supported through the EUR grant NanoX ANR-17-EURE-0009 in the framework of the \textit{``Programme des Investissements d'Avenir''.} 
\end{acknowledgements}

\bibliography{FarDFT}

\end{document}